\documentclass[aps,preprint,noshowpacs,superscriptaddress]{revtex4-2}

\usepackage{graphicx}
\usepackage{dcolumn}
\usepackage{bm}
\usepackage{braket}
\usepackage{appendix}
\usepackage{subfigure}
\usepackage{hyperref}
\usepackage{color}
\usepackage{xcolor}
\usepackage{amsmath}
\usepackage{nccmath}
\usepackage[english]{babel}
\usepackage{svg}
\usepackage{multirow}
\usepackage{tabularx}
\usepackage{array}
\usepackage{blindtext}
\usepackage{titlesec}
\usepackage{soul}
\usepackage{ulem}
\usepackage{amssymb}

\newcolumntype{P}[1]{>{\centering\arraybackslash}p{#1}}

\usepackage{lineno}

\begin{document}

\title{Room-temperature quantum entanglement in a \\ van der Waals material}

\author{Xingyu Gao}
\thanks{These authors contributed equally to this work.}
\affiliation{Department of Physics and Astronomy, Purdue University, West Lafayette, Indiana 47907, USA}

\author{Zhun Ge}
\thanks{These authors contributed equally to this work.}
\affiliation{Department of Physics and Astronomy, Purdue University, West Lafayette, Indiana 47907, USA}

\author{Saakshi Dikshit}%
\thanks{These authors contributed equally to this work.}
\affiliation{Elmore Family School of Electrical and Computer Engineering, Purdue University, West Lafayette, Indiana 47907, USA}
  
\author{Sumukh Vaidya}
\affiliation{Department of Physics and Astronomy, Purdue University, West Lafayette, Indiana 47907, USA}

\author{Peng Ju}
\affiliation{Department of Physics and Astronomy, Purdue University, West Lafayette, Indiana 47907, USA}

\author{Tongcang Li}%
\email{tcli@purdue.edu}
\affiliation{Department of Physics and Astronomy, Purdue University, West Lafayette, Indiana 47907, USA}
\affiliation{Elmore Family School of Electrical and Computer Engineering, Purdue University, West Lafayette, Indiana 47907, USA}
\affiliation{Purdue Quantum Science and Engineering Institute, Purdue University, West Lafayette, Indiana 47907, USA}
\affiliation{Birck Nanotechnology Center, Purdue University, West Lafayette, Indiana 47907, USA}
\date{\today}

\begin{abstract}
 {\normalsize  \bf
 Entanglement is central to quantum science and technology. Atomic defects in two-dimensional (2D) van der Waals (vdW) materials offer exciting prospects for quantum sensing, with spatial resolution reaching 1~nm demonstrated using scanning probe techniques. However, entangling qubits in vdW materials remains elusive. Here we report room-temperature quantum entanglement between an optically addressable electron spin and a strongly coupled $^{13}$C nuclear spin in hexagonal boron nitride (hBN). We extend the electron spin coherence to 38~$\mu$s with dynamical decoupling, and create maximally entangled Bell states with a fidelity up to 0.89. We further use the nuclear spin as a long-lived quantum memory to enhance AC magnetic field sensing via correlation spectroscopy. These results establish entangled spin qubits in hBN as a robust platform for advanced quantum technologies based on 2D materials.
 }
\end{abstract}

\maketitle

Quantum entanglement, a cornerstone of quantum mechanics, is a vital resource for quantum computing, secure communication, and high-precision sensing. Entangled photons \cite{pan2012multiphoton,weissflog2024tunable,lyu2025tunable} have been used to test the foundations of quantum mechanics (e.g. violation of Bell inequalities) \cite{pan2012multiphoton,schirber2022nobel}, realize quantum cryptography \cite{Jennewein2000PhysRevLett.84.4729}, demonstrate quantum computational advantage \cite{zhong2020quantum}, and enhance sensing \cite{xia2023entanglement}. Entanglement between photons and solid-state spin qubits has been utilized to realize quantum networks \cite{pompili2021realization,knaut2024entanglement}. While entanglement among multiple solid-state qubits is critical for scalable quantum information processing, materials that can realize this at room temperature are extremely rare \cite{dutt2007quantum,neumann2008multipartite,bourassa2020entanglement}.

Two-dimensional (2D) van der Waals (vdW) materials have been at the forefront of condensed matter physics and materials science since the discovery of graphene. Recently, atomic defects in 2D vdW materials have emerged as promising platforms for quantum sensing and networking. Remarkably, a scanning probe based on a single atomic defect in a few-layer vdW material achieved 1~nm spatial resolution in imaging moir\'{e} potentials \cite{klein2024imaging,inbar2023quantum}, an order of magnitude improvement in spatial resolution over scanning probes based on diamond nitrogen-vacancy (NV) centers \cite{huxter2023imaging,song2021direct}. Optically active spin defects in vdW material hexagonal boron nitride  (hBN)\cite{gottscholl2020initialization,gao2024single,stern2023quantum,whitefield2025generation} have been used as  ultrathin sensors \cite{durand2023optically,zhou2024sensing} for detecting magnetic fields \cite{huang2022wide,healey2022quantum}, electric fields \cite{gong2023coherent}, temperature \cite{gottscholl2021spin,healey2022quantum}, and etc \cite{vaidya2023quantum}.
These hBN spin defects are ideal for atomic-scale quantum sensing \cite{shen2024proximity}, offer compatibility with device integration\cite{novoselov20162d,healey2022quantum}, and operate across a wide temperature range \cite{gottscholl2021spin}.  Their utility, however, is limited by short electron spin coherence times due to the dense nuclear spin environment. This limitation can be addressed using hybrid electron-nuclear spin systems, which combine fast optical access to electron spins with long coherence of nuclear spins \cite{gao2022nuclear,clua2023isotopic,gong2024isotope,ru2024robust,gao2024single,zaiser2016enhancing,rosskopf2017quantum}.  Achieving this requires precise initialization, control, and readout of both spins. Entanglement between two spin qubits in vdW materials, an essential resource for quantum information science and technologies, has so far remained elusive.

In this work, we demonstrate room-temperature quantum entanglement between an electron spin and a strongly coupled $^{13}$C nuclear spin hosted by a carbon-related spin defect in hBN (Figure \ref{figure1}(a)). These defects exhibit strong hyperfine coupling to individual $^{13}$C nuclear spins, large optically detected magnetic resonance (ODMR) contrast, and narrow linewidths \cite{gao2024single,stern2023quantum,whitefield2025generation}. Using Carr-Purcell-Meiboom-Gill (CPMG) sequences, we extend the electron spin coherence time by two orders of magnitude, from 141 ns to 38 $\mu$s. This is the longest measured coherence time for an electron spin in hBN at room temperature \cite{rizzato2023extending,stern2023quantum,tarkanyi2025understanding}. We generate Bell states between a single electron spin and a single $^{13}$C nuclear spin with a fidelity up to 0.89. We also perform correlation spectroscopy for AC magnetic field sensing. By leveraging the $^{13}$C nuclear spin as a long-lived quantum memory, we extend the correlation time beyond the electron spin's relaxation limit, enabling improved frequency resolution.  These results establish entangled spin qubits in hBN as a viable platform for hybrid quantum devices and open new avenues for quantum sensing and information processing in vdW materials.

\section*{Spin defect with a strongly coupled $^{13}$C nuclear spin register}
\begin{figure*}[ht]
  \centering
 \includegraphics[width=0.94\textwidth]{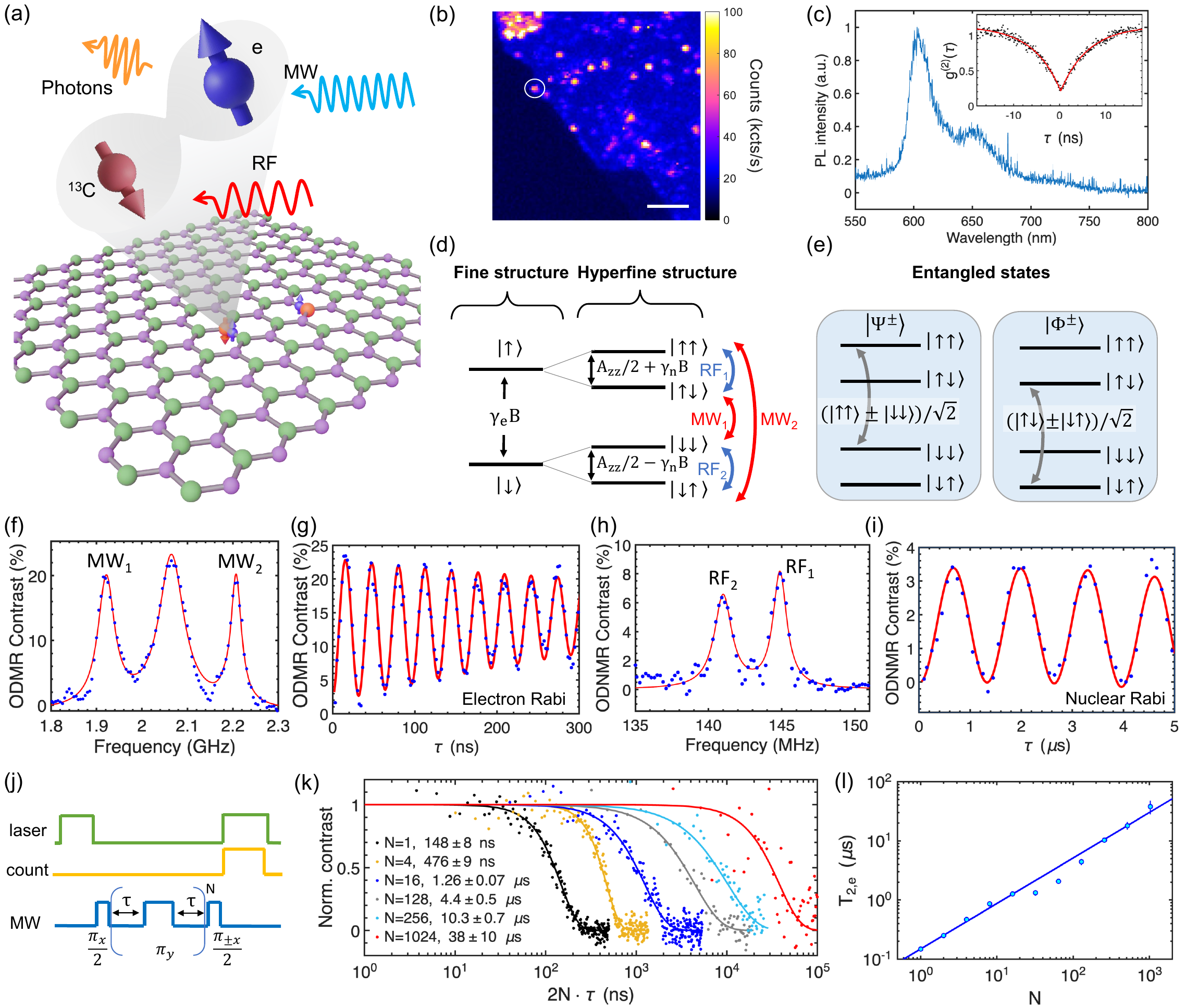}
  \newpage
  \caption{\textbf{Characterization of an electron–nuclear spin system in hBN for entanglement generation.} (a) An hBN spin defect including a $^{13}$C atom. The electron and nuclear spins are controlled using microwave (MW) and radiofrequency (RF) pulses, respectively. (b) Photoluminescence (PL) map of an hBN flake. A white circle marks a spin defect. Scale bar: 5 $\mu$m. (c) PL spectrum of the spin defect under 532 nm laser excitation. Inset: Photon time-correlation measurement.   (d) Simplified energy diagram of an electron spin strongly coupled to a $^{13}$C nuclear spin, with electron spin transitions indicated by red arrows (MW$_1$ and MW$_2$) and nuclear spin transitions labeled by blue arrows (RF$_1$ and RF$_2$). (e) Illustration of four maximally entangled Bell states.  (f) CW ODMR spectrum.  (g)  Rabi oscillation of the electron spin.  (h) ODNMR spectrum of the $^{13}$C nuclear spin.  (i) Rabi oscillation of the $^{13}$C nuclear spin. (j) Dynamical decoupling pulse sequence with $N$ refocusing $\pi$-pulses. (k) Representative measurements for different $N$, showing extended coherence times with increasing pulse number. (l) Extracted electron spin coherence time $T_{2,\mathrm{e}}$ as a function of $N$. All measurements were performed at room temperature in a magnetic field of 73.5 mT.}\label{figure1}
\end{figure*}

We create carbon-related spin defects in hBN via $^{13}$C ion implantation followed by high-temperature thermal annealing \cite{gao2024single}. Confocal photoluminescence (PL) mapping reveals isolated emitters, as shown in Fig. \ref{figure1}(b). The white circle marks a defect used in this work. A representative PL spectrum of the defect displays a peak near 600 nm (Fig. \ref{figure1}(c)). The second-order photon correlation measurement yields $g^{(2)}(0)=0.21<0.5$, confirming that the emission originates from a single defect complex.

The defect can be described as a spin pair system, consisting of two electron spins localized at closely spaced point defects separated by only a few nanometres (Figure \ref{figure1}(a)) \cite{robertson2024universal,gao2024single,li2025quantum}. For this defect, a strong hyperfine interaction is established between one of the electron spins and the nuclear spin of the $^{13}$C atom in the defect, while the second electron spin has a much weaker hyperfine coupling to nearby nuclear spins \cite{gao2024single}. This allows us to isolate and focus on a spin subsystem composed of a single electron spin and its strongly coupled $^{13}$C nuclear spin, forming a well-defined four-level, two-qubit system, as shown in Figure~\ref{figure1}(d). The strong hyperfine coupling enables selective control of individual transitions using microwave (MW) and radiofrequency (RF) pulses.

This simplified four-level two-qubit system with a single electron spin-$\frac{1}{2}$ and a single $^{13}$C nuclear spin-$\frac{1}{2}$ in Fig.~\ref{figure1}(d) can be described by the Hamiltonian:
\begin{equation}
    H = \gamma_e \mathbf{B}_0 \mathbf{S} + \gamma_n \mathbf{B}_0 \mathbf{I} + \mathbf{S}\cdot \mathbf{A} \cdot  \mathbf{I},
\end{equation}
where $\mathbf{S}$ and $\mathbf{I}$ are the spin-1/2 operators for electron and nuclear spins, respectively. $\gamma_e$ and $\gamma_n$ are the gyromagnetic ratios of the electron spin and the $^{13}$C nuclear spin. $\mathbf{A}$ represents the hyperfine coupling tensor. In this simplified picture, the second electron spin can be ignored because its transition is well separated and remains unaddressed during the resonant driving of the target electron-nuclear spin subspace. 
Within this two-qubit system, the four maximally entangled Bell states can be defined as $\ket{\Psi^\pm} = (\ket{\uparrow\uparrow} \pm \ket{\downarrow\downarrow})/\sqrt{2}$ and $\ket{\Phi^\pm} = (\ket{\uparrow\downarrow} \pm \ket{\downarrow\uparrow})/\sqrt{2}$ (Figure \ref{figure1}(e)).

The continuous-wave (CW) ODMR spectrum reveals two well-separated resonances with a splitting of approximately 290~MHz (Figure \ref{figure1}(f)), attributed to one member of the defect pair, likely a C$_{B}$O$_{N}$ defect\cite{gao2024single,guo2023coherent}. A third, central resonance is also observed and corresponds to the second electron spin in the pair. For most defects, this central transition is well resolved and separated from the two 290~MHz hyperfine-split sidebands (see supplemental Fig. S1). 
We can drive individual electron spin transition using selective MW pulses (MW$_1$ or MW$_2$), which results in Rabi oscillation as shown in Figure \ref{figure1}(g). We can further identify the nuclear magnetic resonance (NMR) frequency of the strongly coupled $^{13}$C nucleus using optically detected nuclear magnetic resonance (ODNMR). The ODNMR spectrum reveals two well-resolved resonance peaks at 141 MHz and 145 MHz (Figure \ref{figure1}(h)), corresponding to nuclear spin transitions with different electron spin states. Under a 1~W RF drive, we observe nuclear Rabi oscillation with a frequency of approximately 0.8 MHz (Figure \ref{figure1}(i)). This fast nuclear spin operation is enabled by an effective enhancement of the nuclear gyromagnetic ratio due to the strong hyperfine interaction in the relatively low magnetic field regime (73.5 mT) \cite{gao2022nuclear}.

To extend the coherence of the electron spin, we implement the CPMG dynamical decoupling protocol, which consists of a series of evenly spaced microwave $\pi$-pulses between two $\pi/2$ pulses (Figure \ref{figure1}(j)). This sequence suppresses low-frequency noise and environmental decoherence, allowing the spin coherence to be preserved for longer durations compared to a single Hahn echo. For reference, the Hahn echo measurement yields a coherence time of $T_{2,\mathrm{echo}} = 148 \pm 8$ ns. As shown in Figure \ref{figure1}(k), increasing the number of refocusing pulses leads to systematically longer coherence times, reaching $T_{2,e} = 38 \pm 10\,\mu\text{s}$ with CPMG-1024. Figure \ref{figure1}(l) shows the extracted coherence time follows a power-law scaling, $T_{2,e} \propto N^\beta$, with a fitted exponent of $\beta = 0.77 \pm 0.09$. This behavior is consistent with decoherence dominated by slowly fluctuating environmental noise and highlights the efficacy of CPMG control in extending spin coherence in 2D spin systems.

\section*{Entangling an electron spin and a nuclear spin in hBN}  
\begin{figure*}[ht]
  \centering
  \includegraphics[width=1\textwidth]{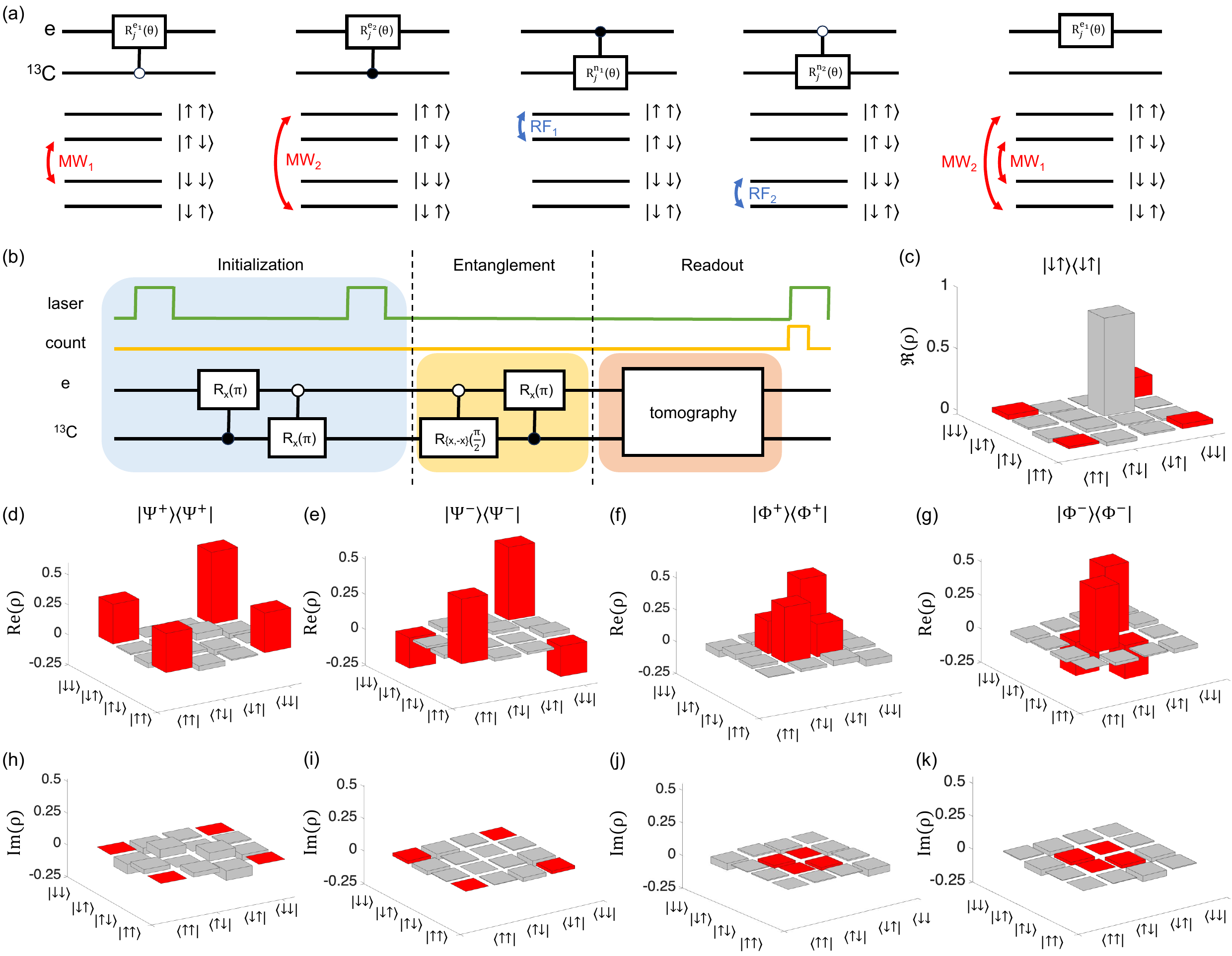}
  \caption{\textbf{Entanglement generation between an electron spin and a nuclear spin in hBN.}  (a) Schematic illustration of non-local and local control gates. Non-local spin gates are implemented by driving a single hyperfine-resolved transition, while local spin gates are realized by simultaneously driving two transitions. j (= $\pm$ x, or $\pm$ y) refers to the global phase of the gates. (b)  Experimental sequence for generating and characterizing the entangled state. The sequence includes initialization of the electron and nuclear spins, a series of conditional gate operations to create entanglement, and quantum state tomography using electron spin rotations and fluorescence readout. (c) Real part of the reconstructed density matrices of the spin system after spin initialization. (d)-(g) The real part of the density matrices after entanglement generation to create the (d) $|\Psi^+\rangle$, (e) $|\Psi^-\rangle$, (f) $|\Phi^+\rangle$, and (g) $|\Phi^-\rangle$ Bell states. (h)-(k) The imaginary part of the density matrices after entanglement generation to create the (h) $|\Psi^+\rangle$, (i) $|\Psi^-\rangle$, (j) $|\Phi^+\rangle$, and (k) $|\Phi^-\rangle$ Bell states. }\label{figure2}
\end{figure*}

Within the simplified electron-nuclear spin subspace, we can introduce two-qubit gate operations for systematic spin control and entanglement generation between the electron and nuclear spins. As illustrated in Fig.~\ref{figure2}(a), nonlocal gates, where the operation on one qubit is conditional on the state of the other, are realized by selectively driving individual hyperfine-resolved transitions. For the electron spin, this can be realized by driving individual electron spin transition using MW$_1$ or MW$_2$ (Fig. \ref{figure1}(g)).
Similarly, selective RF excitation on RF$_1$ or RF$_2$ enables coherent control of the nuclear spin (Fig.~\ref{figure1}(i)).

In contrast, local gates for the electron spin, operations that address
the electron spin irrespective of the state of the nuclear spin, are implemented by applying 
MW$_1$ and MW$_2$ drives at both transition frequencies simultaneously (Fig.~\ref{figure2}(a)). The local gate requires two MW tones separated by approximately 290 MHz. In practice, this frequency separation results in slightly unequal Rabi frequencies due to variations in MW transmission and device response across the two frequencies. If not corrected, this mismatch manifests itself as a beating in the Rabi oscillation signal, arising from the interference between the two slightly different Rabi drives. To eliminate this effect, we carefully calibrate the amplitude and phase of the dual-tone MW drive to ensure matched Rabi frequencies, thereby suppressing the beating and recovering clean Rabi oscillations (Supplemental Fig. S2).

The electron and nuclear spin gate operations described above enable the preparation of entangled states and the implementation of quantum state tomography (Supplemental Fig. S7) to reconstruct the density matrix and evaluate the resulting states. Figure \ref{figure2}(b) outlines the experimental protocol for generating and characterizing electron–nuclear spin entangled states.  The sequence begins with optical initialization of the electron spin, followed by a SWAP gate, comprising a c-NOT$_e$ gate and a c-NOT$_n$ gate, to transfer spin polarization from the electron to the nuclear spin. A subsequent laser pulse reinitializes the electron spin, resulting in the joint state $|\downarrow_e \uparrow_n\rangle$. The reconstructed density matrix from quantum state tomography confirms strong population in this target state, indicating successful initialization (Fig.~\ref{figure2}(c)).

To generate entanglement, we first apply a nonlocal gate to the nuclear spin, coherently transforming the system into the superposition state $\left(|\downarrow_e \uparrow_n\rangle \pm |\downarrow_e \downarrow_n\rangle\right)/\sqrt{2}$. This is followed by a nonlocal gate on the electron spin, which maps the nuclear spin superposition onto the electron spin, producing the entangled Bell state $|\Psi^\pm\rangle$ = $\left(|\uparrow_e \uparrow_n\rangle \pm |\downarrow_e \downarrow_n\rangle\right)/\sqrt{2}$. Figures~\ref{figure2}(d) and \ref{figure2}(e) show the real parts of the reconstructed density matrices corresponding to the prepared $|\Psi^+\rangle$ and $|\Psi^-\rangle$ states, respectively. We also prepare the other two Bell states, $|\Phi^\pm\rangle$ = $\left(|\uparrow_e \downarrow_n\rangle \pm |\downarrow_e \uparrow_n\rangle\right)/\sqrt{2}$ as shown in Figures~\ref{figure2}(f) and \ref{figure2}(g). The imaginary components of the reconstructed density matrices for all four Bell states are provided in Figures~\ref{figure2}(h), (i), (j), (k). 
The fidelity of the reconstructed two-qubit density matrix $\rho$  with respect to an ideal Bell state $ \rho_{\text{ideal}} = |\psi\rangle \langle\psi|$ can be evaluated by 
$F =  \mathrm{Tr} \left( \sqrt{ \sqrt{\rho_{\text{ideal}}} \, \rho \, \sqrt{\rho_{\text{ideal}}} } \right)$ \cite{Hu2024QuantumRegister}.
By comparing the experimentally reconstructed density matrix for the $|\Psi^+\rangle$ state with the ideal Bell state $\left(|\uparrow_e \uparrow_n\rangle + |\downarrow_e \downarrow_n\rangle\right)/\sqrt{2}$, we extract an entanglement fidelity of up to 0.89 under the pseudo-pure approximation \cite{simmons2011entanglement}.

\section*{Nuclear spin memory for enhanced sensing}
The ancilla nuclear spin can function as a long-live quantum memory, extending the coherence of the electron spin sensor through state transfer protocols. By applying a sequence of conditional operation, specifically a c-NOT$_e$ gate  followed by a c-NOT$_n$ gate, the electron spin is coherently mapped onto the nuclear spin state, allowing temporary storage of quantum phase information accumulated by the electron spin. The reverse sequence retrieves the stored state and restores it to the electron spin for subsequent manipulation or readout.

\begin{figure*}[htbp]
  \centering
  \includegraphics[width=1\textwidth]{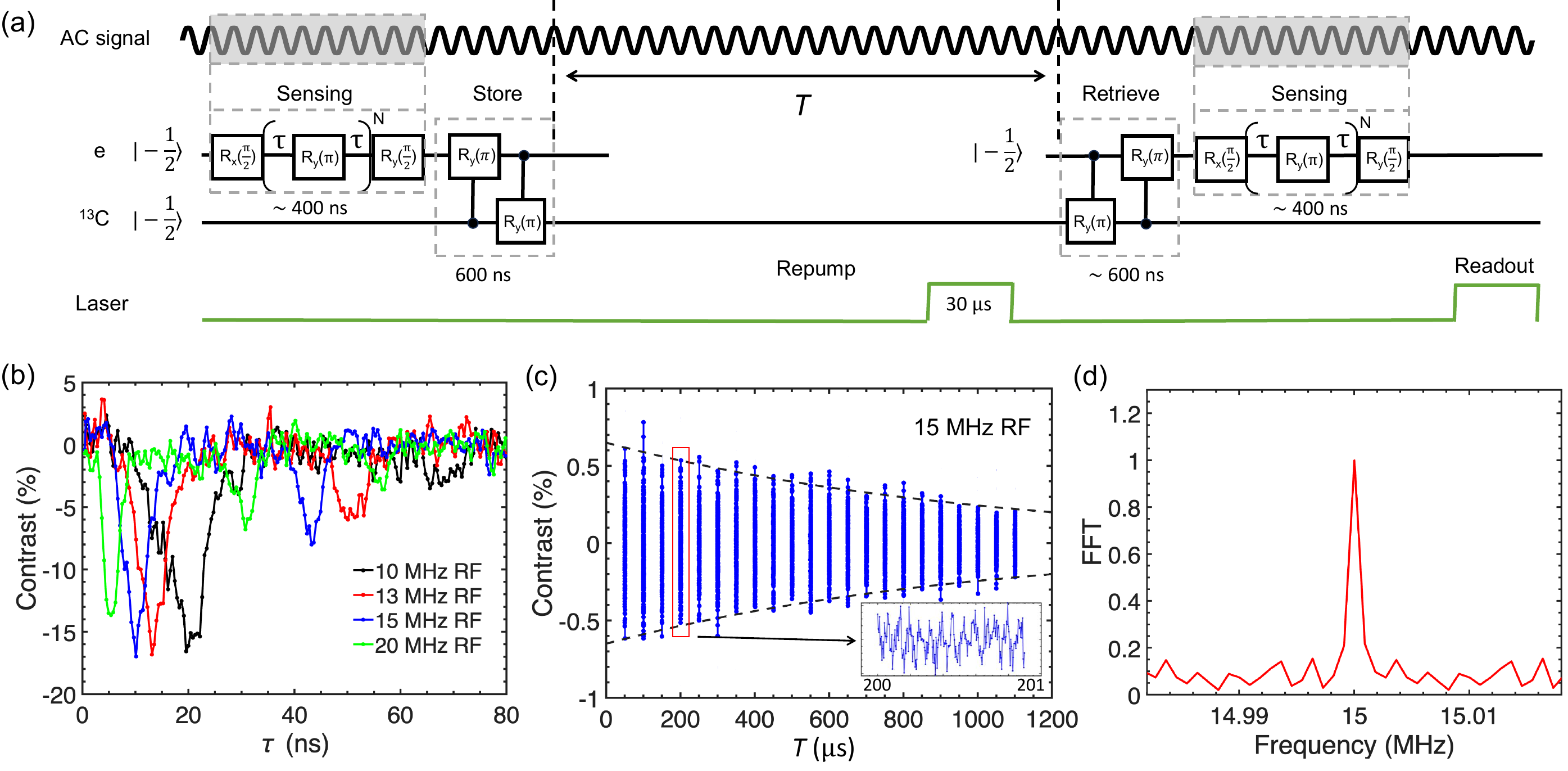}
  \caption{\textbf{Enhancing quantum sensing with a nuclear spin quantum memory in hBN.}  (a) Schematic of the correlation measurement protocol using the nuclear spin as a quantum memory. The electron spin acquires a phase through a CPMG pulse sequence tuned to the target AC field, which is then stored in the nuclear spin via a SWAP gate. After a variable delay time $T$, the state is retrieved, and a second CPMG sequence accumulates an additional phase for correlation readout. (b) CPMG-8 measurements in the presence of externally applied RF field, from 10 MHz to 20 MHz. A clear dip is observed when $\tau$ matches the resonance conditions. (c) Correlation measurement with nuclear spin memory. The information is preserved over a long delay beyond electron spin relaxation time and is ultimately limited by the nuclear spin relaxation time. (d) Fourier spectra of the signals in (c), showing a frequency measurement precision of 1 kHz using nuclear spin memory. }\label{figure3}
\end{figure*}

We harness this quantum memory functionality to implement a correlation spectroscopy protocol for detecting unsynchronized AC magnetic fields. The correlation spectroscopy does not require phase synchronization between the target signal and the sensing sequence, making it suited for sensing fluctuating or ambient signals \cite{laraoui2013high}. As shown in Figure \ref{figure3}(a), the procedure begins by initializing the electron and nuclear spins, followed by applying a CPMG sequence on the electron spin tuned near the resonance of the target AC field, $ V(t) = V_0 \cos(2\pi f_{\text{ac}} t)$. The CPMG sequence imprints a phase $\phi_1$ on the electron spin coherence, dependent on the amplitude and frequency of the AC field. The quantum phase is transferred to the population with an electron spin $\pi/2$ gate and then stored in the nuclear spin, which serves as a storage unit during an idle period of duration $T$.

After this delay, the spin state is retrieved and subjected to a second CPMG sequence, during which an additional phase $\phi_2$ is accumulated under the continued influence of the AC field. By measuring the final electron spin population and repeating the protocol over multiple trials with randomized field phases, we extract the averaged correlation signal. The resulting population oscillates with the delay time $T$ as \cite{rosskopf2017quantum}
\begin{equation}
    p(T) = \frac{1 - \langle \sin \phi_1 \sin \phi_2 \rangle}{2} \approx \frac{1 - p_0 \cos(2\pi f_{\text{ac}} T)}{2}
\end{equation}
revealing spectral information about the field even when it is not synchronized with the experimental control sequence. Since the delay time $T$ is limited by the nuclear spin relaxation time $T_{1,n}$ rather than the shorter electron spin relaxation time $T_{1,e}$, this memory-assisted scheme enhances both sensitivity and spectral resolution by enabling phase correlation measurements beyond the intrinsic limit of the electron spin.

To determine the resonance condition for interpulse delay time $\tau$ in each CPMG sensing module, we begin with standard CPMG measurements by sweeping $\tau$. As shown in Figure~\ref{figure3}(b), in the presence of different RF fields with frequencies ranging from 10 to 20 MHz, a pronounced dip in the spin contrast is observed when the CPMG sequence is resonant with the applied RF signal, satisfying satisfying the condition $\tau = k/(4f_{\text{ac}})$ (k = 1,3,5, ...). This demonstrates the capability of the spin sensor to detect AC magnetic fields with megahertz-level frequency resolution. The AC magnetic field sensitivity is approximated 10~$\mu$T$\sqrt{\rm Hz}$ for a single spin (Supplemental Section VI). The spectral response of the CPMG sequence effectively acts as a bandpass filter. However, due to the finite duration of the $\pi$ pulses, the effective sensing interval deviates slightly from the nominal value, resulting in a systematic offset from the expected $\tau = k/(4f_{\text{ac}})$ \cite{zaiser2016enhancing}.

This offset can be corrected and precisely calibrated using correlation measurements. To evaluate this method, we detect a 15 MHz RF field with and without employing the nuclear spin memory. As shown in Figure~\ref{figure3}(c), when the nuclear spin memory is used, the signal persists for up to approximately 1 ms, well beyond the electron spin relaxation time. Fourier analysis of this extended oscillation yields a spectral linewidth of 0.7$\pm$0.2 kHz (Figure~\ref{figure3}(d)). In contrast, when the memory protocol is omitted (by removing the store, repump, and retrieve steps), the oscillation decays much more rapidly, with a time constant of approximately 130 $\mu$s, limited by the electron spin relaxation at room temperature (Supplemental Figure S10).

\section*{Conclusion} 
In summary, we have demonstrated room-temperature quantum entanglement between an electron spin and a strongly coupled $^{13}$C nuclear spin in a van der Waals material. Through high-fidelity gate operations and quantum state tomography, we verified entangled state generation. We also used CMPG dynamical decoupling protocol to extend the electron spin coherence time and to sense AC magnetic fields. Furthermore, we leveraged the nuclear spin as a quantum memory to extend the phase correlation time in a sensing protocol, enhancing both sensitivity and spectral resolution for detecting weak AC magnetic fields. These results highlight the potential of carbon-related defects in hBN as a robust platform for hybrid spin quantum devices and open new avenues for quantum sensing and information processing in atomically thin materials. These 2D quantum sensors are particularly promising for probing 2D magnetism \cite{huang2022wide,healey2022quantum}, moir\'{e} physics \cite{klein2024imaging,inbar2023quantum}, and many-body entanglement in emerging quantum materials \cite{zeng2019quantum}. 

\newpage

\clearpage
\newpage


\end{document}